\documentclass{emulateapj}

\shorttitle{The highly unusual composition of the  Hercules dSph}
\shortauthors{Koch et al.}
\slugcomment{Accepted for publication in ApJ Letters}

\begin{document}

\title{The Highly Unusual Chemical Composition of the Hercules Dwarf Spheroidal 
Galaxy\altaffilmark{1}}

\author{Andreas Koch\altaffilmark{2}, Andrew McWilliam\altaffilmark{2}, 
Eva K. Grebel\altaffilmark{3}, Daniel B. Zucker\altaffilmark{4}, \& Vasily Belokurov\altaffilmark{4}}
\altaffiltext{1}{This paper includes data gathered with the 6.5 meter Magellan Telescopes located at 
Las Campanas Observatory, Chile.}
\altaffiltext{2}{Observatories of the Carnegie Institution Washington, Pasadena, CA, USA}
\altaffiltext{3}{Astronomisches Rechen-Institut, Zentrum f\"ur Astronomie der 
Universit\"at  Heidelberg, Germany}
\altaffiltext{4}{Institute of Astronomy, Madingley Road, Cambridge, UK}
\email{akoch@ociw.edu}

\begin{abstract}
We report on the abundance analysis of two red giants in the 
 faint Hercules dwarf spheroidal (dSph) galaxy. 
These stars show a remarkable deficiency in the neutron-capture elements,  
while the hydrostatic $\alpha$-elements (O, Mg) are strongly enhanced. 
Our data indicate [Ba/Fe] and [Mg/Fe] abundance ratios of $\la$$-2$ dex and $\sim$+0.8 dex, respectively, 
with essentially no detection of other n-capture elements.  
In contrast to the only other dSph star with similar  abundance patterns, 
Dra~119, which has a very low metallicity at [Fe/H]=$-$2.95 dex, our objects, at 
[Fe/H]$\sim$$-2.0$ dex, are only moderately  metal poor. 
The measured ratio of hydrostatic/explosive $\alpha$-elements indicates
that high-mass ($\sim35$~M$_{\sun}$) Type II supernovae progenitors are the
main, if not only, contributors to the enrichment of this galaxy.  This
suggests that star formation and chemical enrichment in the ultrafaint
dSphs proceeds stochastically and inhomogeneously on small scales,  or that the IMF was strongly
skewed to high mass stars. 
The neutron capture deficiencies and the [Co/Fe] and [Cr/Fe] abundance ratios
in our stars are similar to those in the extremely low metallicity Galactic 
halo.  This suggests that either our stars are composed mainly of the ejecta from the first, massive,   
population~III stars (but at moderately high [Fe/H]), or that SN ejecta
in the Hercules galaxy were diluted with $\sim$30 times less hydrogen than
typical for extreme metal-poor stars.  
\end{abstract}


\section{Introduction}
Chemical abundance patterns in stars reflect the star formation (SF) histories of stellar systems:  
whatever mechanisms contributed to the enrichment of the interstellar medium
will leave unique imprints in the resulting  
chemical abundance ratios. These  allow one not only to trace the predominant 
modes of SF and reconstruct galactic enrichment histories, but also to place important 
constraints on the nucleosynthetic production sites of the elements. 
In particular, the $\alpha$-elements (e.g., O, Mg, Si, Ca, Ti), produced mainly by massive stars which 
end as Type II supernovae (SNe) on short time scales, and the neutron-capture elements (e.g. Y, Sr, Ba, La)  from AGB
stars, probe rapid and slow timescales, respectively.

Knowledge about chemical abundances and the evolutionary histories of 
the more luminous dwarf spheroidal (dSph) galaxies is constantly growing (e.g., Shetrone et al. 2003; 
Venn et al. 2004; Koch et al. 2008). However,   
only little is known about the metal-enrichment of the 
ultra-faint, presumably low-mass, systems recently discovered in the Sloan Digital Sky Survey 
(SDSS;  Zucker et al. 2006a,b; Belokurov et al. 2006, 2007, 2008; Irwin et al. 2007; Walsh et al. 2007).  
Current low-resolution spectroscopy and photometric data
indicate that these galaxies are predominantly metal poor 
(e.g., Mu\~noz et al. 2006; Belokurov et al. 2007; Kirby et al. 2008). 

The discovery of only a couple of chemically peculiar stars in Galactic dSphs (Fulbright et al. 2004; 
Sadakane et al. 2004; Sbordone et al. 2007; Koch et al. 2008) and their similarity to a number of anomalous halo objects (Carney et al. 1997; 
Ivans et al.  2003) poses important questions for the general understanding of 
galaxy formation and chemical evolution: how common are chemically peculiar stars in the dSphs? 
What were the predominant enrichment processes that led to their abundance patterns? 
Are peculiar stars found in the Galactic halo related to those in dSphs? 

Hercules (hereafter Her) is one of the faintest ($M_V=-6.6$ mag) 
and among the very metal poor ($\sim -2.3$ dex) dSph candidates 
discovered in the SDSS (Belokurov et al. 2007). 
Its {\em stellar} mass is estimated to be 4--7$\times 10^4$~M$_{\sun}$ 
(Martin et al. 2008). 
Its extended morphology (Coleman et al. 2007) and a broad RGB 
render it an intrinsically peculiar galaxy that may have experienced interactions with  the Milky Way. 
An investigation of its chemical properties is thus a natural endeavor. 

In this Letter we present the discovery of unusual chemical abundance patterns in two red giants
in Hercules. The elemental abundances were identified in the course of 
a broader observational program dedicated to a high-resolution study of Hercules, and the 
comprehensive  abundance analysis of the full sample will be presented elsewhere 
(Koch et al. 2008, in prep.) 

\section{Observations and Abundance Analysis}
Targets were selected from the  Sixth Data Release of the SDSS  (Adelman-McCarthy et al. 2008). 
There is only a limited number of brighter red giants present in Her that are observable within 
reasonable integration times (V$\la$19.2 mag)  
and non-members were rejected from the sample in real time during the observing run, 
based on their highly discrepant velocities (cf. Simon \& Geha 2007).  
Here we present the results  for two of our red giants, which we label Her-2 and Her-3 according 
to their brightness (V=18.7 and 19.0 mag). The observations were carried out over two nights on 
2007 July 10--11 with the Magellan Inamori Kyocera Echelle (MIKE) 
spectrograph at the 6.5-m Magellan2/Clay Telescope. 
A slit width of 1$\arcsec$ and a CCD binning of 2$\times$1 pixel resulted in 
a spectral resolution of 20000, with a full spectral coverage using the blue and red echelles  
 of 3350-9400\AA. 
The  stars were observed for a total of four  and six hours, respectively. 
Our data were reduced with the standard 
MIKE pipeline reduction package of Kelson (2003). 
The  signal-to-noise (S/N) ratio of our spectra is 32 per pixel at $\sim$6500\AA, 
but  it falls below 10 for the bluer orders. 
The  radial velocities of the two target stars confirm their membership in Her 
(Simon \& Geha 2007), but we note that Her-3  is probably a binary, based on its 
radial velocity curve (D. Ad\'en 2008, private 
communication), although there is no secondary flux detected from the 
 companion.
Therefore, the companion is most likely very faint and  will 
not affect the present analysis. 

The chemical abundances were determined from an equivalent width (EW) analysis, where 
we employed $gf$-values from the literature.  
Typically, 15--25 \ion{Fe}{1} lines of sufficient strength could be measured. 
For most of the other species, only a few absorption features were distinguishable (see 
Table~1), which, coupled with the slightly low S/N led to large scatter in some abundance ratios. 
Many absorption lines of the heavy elements are too weak to be reliably 
measured in our spectra, yielding only upper abundance limits.

Stellar abundances 
were computed with the program MOOG (Sneden 1973) using  
stellar atmospheres generated from the 
Kurucz LTE models\footnote{\url{http://kurucz.harvard.edu}}. 
Given the strong enhancements in the $\alpha$-elements O and Mg in our Her stars 
(Sect.~3), we chose to use the $\alpha$-enhanced Kurucz model atmospheres, AODFNEW, 
in our analysis.  
Furthermore, we adopted photometric ($V$$-$$I$) effective temperatures 
for our stellar atmospheres, where colors were transformed from 
the SDSS g,r,i system using the prescriptions of Jordi et al. (2006). 
A reddening of E(B$-$V)= 0.06 (Schlegel 1998) and the according extinction laws of 
Rieke \& Lebovsky (1985) were adopted throughout this work. 
We note that the photometric values agree well with the temperatures derived 
from excitation equilibrium.  The physical  gravities were derived from the photometric data, 
adopting a distance to Her of 140 kpc (Belokurov et al. 2007) and a stellar mass of 0.8 M$_{\sun}$, 
representative of the red giants (see also Koch \& McWilliam 2008).
The Fe abundances from  the ionized and neutral lines agree marginally within the uncertainties so that 
ionization equilibrium holds with the adopted surface gravities. 
As a result, we derive stellar atmosphere parameters (T$_{\rm eff}$, log\,$g$, $\xi$) of 
(4270 K, 0.69, 2.5 km\,s$^{-1}$) for Her-2 
and (4340 K, 0.91, 2.5 km\,s$^{-1}$) for Her-3, with typical uncertainties of 
($\pm$50 K, $\pm$0.15, $\pm$0.1  km\,s$^{-1}$) . 
Our adopted microturbulent velocity values are higher than 
typical red giant stars with similar parameters.  This is likely an artifact of the relatively low S/N 
of our spectra, due to asymmetry in abundance errors on strong
lines (Magain 1984); however, we stress that these high 
microturbulent velocity values are appropriate for our spectra. 
Hyperfine splitting was included for the even-Z elements Sc, Mn, Co, and Ba, but is usually 
negligible for the Na lines we used in our stars. 
\section{Abundance results}
The final chemical abundance ratios derived for the red giants are listed 
in Table~1, relative to \ion{Fe}{1} for the neutral species and to \ion{Fe}{2} for ionized 
species and for 
the [\ion{O}{1}] lines. Solar abundances were taken from Asplund et al. (2005).  
Listed in  Table ~1 are 1\,$\sigma$ scatter of our 
measurements from individual lines and the numbers of lines employed to derive these
abundances. The above uncertainties on the atmospheric parameters translate into 
typical total systematic errors of less than 0.1 dex, based on the formalism 
of Koch \& McWilliam (2008; their Table 8).
Details on the full data set  and a thorough treatment of all 
systematic errors will be presented in a forthcoming paper. 
\begin{deluxetable}{rrrrrrrrr}
\tabletypesize{\scriptsize}
\tablecaption{Abundance results}
\tablewidth{0pt}
\tablehead {\colhead{} & \multicolumn{4}{c}{Her-2} &  \multicolumn{4}{c}{Her-3} \\  
\cline{2-4} \cline{6-9}                                       
\raisebox{1.5ex}[-1.5ex]{Ion} & \colhead{{[}X/Fe{]}} & \colhead{$\sigma$}& \colhead{N} &
\colhead{} & \colhead{{[}X/Fe{]}} & \colhead{$\sigma$}& \colhead{N}}
\startdata
{[}Fe/H{]}$_{\rm CaT}$\tablenotemark{a} & $-$2.39 & 0.13 & \nodata & & $-$1.72 & 0.15 & \nodata &  \\
{[}Fe\,I/H{]}  & $-$2.02 & 0.20 & 24 & & $-$2.04 & 0.43 & 15  \\
{[}Fe\,II/H{]} & $-$1.78 & 0.21 &   3 & & $-$1.84 & 0.41 &   2  \\
{[}O\,I{]}  &    1.12 &    0.18 &  2 & & $<$0.89 &    0.08 & 2   \\
{Na\,I}  &    0.78 &    0.11 &  2 & &    0.62 &    0.09 & 2   \\
{Mg\,I}  &    0.81 &    \nodata &  1 & &    0.77 &  \nodata  & 1   \\
{Al\,I}  &   $<$0.75 &  \nodata &  1 & & \nodata & \nodata & \nodata   \\
{Si\,I}  & $<$0.58 & \nodata &  1 & &    $<$0.63 & \nodata & 1   \\
{Ca\,I}  & $-$0.13 &    0.05 &  7 & &    0.19 &    0.09 & 3   \\
{Sc\,II} &    0.04 &    0.17 &  3 & &    0.26 &    0.20 & 2   \\
{Ti\,I}  &    0.17 &    0.23 &  9 & &    0.33 &    0.13 & 3   \\
{Cr\,I}  &    $-$0.05 &    0.32 &  3 & &    $-$0.14 &    0.04 & 2   \\
{Mn\,I}  &    0.23 &    0.08 &  2 & & $-$0.06 & 0.33 & 3  \\
{Co\,I}  &    0.53 &    0.10 &  2 & &    0.42 & \nodata & 1   \\
{Ni\,I}  &    0.30 &    0.10 & 10 & &    0.12 &    0.15 & 5   \\
{Ba\,II} & $<$$-$2.14 & 0.06 &  2 & & $<$$-$2.10 & 0.06 & 2 
\enddata
\tablenotetext{a}{Metallicity estimate based on the calcium triplet calibration of Rutledge 
et al. (1997a,b), on the metallicity scale of Carretta \& Gratton (1997).}
\end{deluxetable}

\subsection{Iron}
At an [Fe/H] of $-2.02$ and $-2.04$  dex, both stars are slightly more metal rich than the sample mean 
found from comparison with 
globular cluster fiducials (Belokurov et al. 2007) and from low-resolution 
data, which  yield a mean metallicity of $-2.3$ or $-2.5$ dex   
(Simon \& Geha 2007; Kirby et al. 2008). 
The latter work finds an intrinsic metallicity spread of at least 0.5 dex (r.m.s.), which would indicate that our stars
are well within the Her metal-rich tail.  In particular, almost all of the ultrafaint dSphs studied to 
date show broad metallicity ranges of this order of magnitude. 
In general, a zero-point difference between the CaT ``metallicities''
and our [Fe/H] values can be expected due to the low [Ca/Fe] ($-$0.15) found here.   
Na\"ively, we expect a zero-point correction to the CaT metallicities based on halo [Ca/Fe] calibrations 
of $+$0.50 dex (see also Koch et al. 2008). 
This would suggest [Fe/H]=$-$1.8 from correcting the Simon \& Geha (2007)
 metallicities.  We can check this result using our MIKE spectra, which 
contain the CaT lines; accordingly, we find CaT-based metallicities of [Fe/H]$_{\rm CaT}= -$2.39 and $-$1.72 for Her-2 and 3.  
Moreover, our high-resolution [Fe/H] for Her-2 agrees well to within the uncertainties with the  
low-dispersion value derived by Kirby et al. (2008) and with the value based on  Str\"omgren photometry
(D. Ad\'en; private communication). 
\subsection{Alpha-  and light elements}
As a glance at Table~1 shows, 
the [Ca/Fe] and [Ti/Fe] ratios of the Her stars are at most slightly to moderately enhanced, 
which is compatible 
with the abundance ratios found in the more luminous dSphs (Fig.~1; Shetrone et al. 2001,2003; 
Venn et al. 2004; Sadakane et al.\ 2004; Sbordone et al.\ 2007; Koch et al. 2008). 
\begin{figure}[ht]
\begin{center}
\vspace{0.5cm}
\includegraphics[angle=0,width=1\hsize]{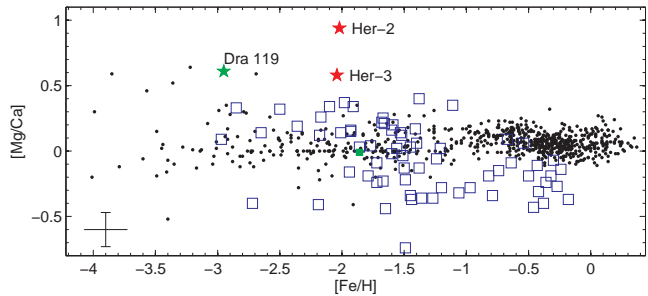}
\end{center}
\caption{Hydrostatic (Mg) to explosive (Ca) $\alpha$-element abundance ratio  
for our two Her giants (red stars) in comparison with Galactic disk, bulge and halo stars (black dots; Venn et al. 2004, and references therein) 
and stars in the luminous Galactic dsphs (blue squares). Highlighted  as green symbols are the 
chemically peculiar 
dSph star Dra~119 (Fulbright et al. 2004) and the $\alpha$-depleted halo star BD+80$\degr$~245 (Carney et al. 1997). Typical 1$\sigma$ 
random error bars for the Her stars are indicated toward the lower left of each panel.}
\end{figure}

At [Si/Fe]$\la0.6$ dex, the Her stars have abundance ratios similar to  other 
metal poor dSph stars from the literature (yet consistent with the typical Galactic halo pattern). 
One has to keep in mind, however, that these values are upper limits based on only one line. 
Furthermore, the production of Si in massive stars, which are required to 
contribute to peculiar abundance patterns in metal-poor stars,  is not yet fully understood (e.g., Cohen et 
al. 2007). 

It is striking that both the hydrostatic $\alpha$-elements Mg and O are strongly 
enhanced by a factor of about 
10 relative to the Solar abundance ratios found in the 
majority of the Local Group dSphs. 
The light element Na is
strongly enhanced in the Her stars, with typical [Na/Fe] ratios of 0.6--0.8 dex. 
For stars with similar atmosphere parameters to Her-2 and 3, the 
calculations of Takeda et al. (2003) show downward non-LTE corrections that
become more severe with increasing EW for our Na~I lines at 8183 and 8195\AA .
Given the large EWs of our 8183/8195 \AA\ Na I lines, near 160--190 m\AA ,
the Takeda et al. results suggest abundance corrections that reduce
our LTE values by 0.3 to 0.4 dex.  Thus, if non-LTE effects on Fe are 
ignored, the [Na/Fe] ratios in our Her stars are probably closer to 
$+$0.3 dex, rather than $\sim$0.7 dex indicated in Table~1.
The weak Al lines are 
 not detectable in Her-3, but Al appears to be enhanced in Her-2.  
This is well consistent with Al and Na production in massive SNe~II, similar to the $\alpha$-elements 
Mg and O  (e.g., Woosley \& Weaver 1995).
\subsection{Neutron-capture elements}
Fig.~2 shows the spectral regions around the \ion{Ba}{2} line at  6496\AA. 
For comparison, we overplot 
a spectrum of a red giant in the Carina dSph, with
 similar atmospheric parameters as the Her stars and with  
a low [Fe/H] of $-$2.72 dex (Koch et al. 2008). The Carina giant has
[Ba/Fe] of $\sim -0.63$ dex and is thus Ba-depleted like the metal poor halo stars. 
Yet, a weak Ba absorption feature is discernible. 
Neither of the Her stars, however,  shows any Ba absorption 
above the noise level. 
As a secondary test we synthesized the spectral regions around the Ba 6141 and 6496 lines 
and verified that essentially all of the visible absorption is due to noise and/or a blend with an Fe line.  
As a result, the [Ba/Fe] ratio is compatible with upper limits from the spectral S/N of $\sim -2$ dex in both stars (Fig.~3). 
\begin{figure}[tb]
\begin{center}
\includegraphics[angle=0,width=1\hsize]{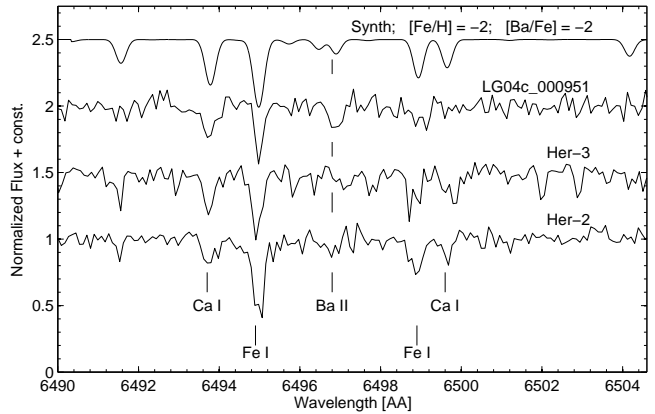}
\end{center}
\caption{MIKE spectra of the two Hercules stars 
centered at the \ion{Ba}{2} 6496 line. Selected absorption lines are 
labelled. Shown for comparison are a UVES spectrum of a metal poor ([Fe/H]=$-2.72$) star in the Carina dSph that 
has stellar parameters similar to the Her stars and shows a moderate Ba depletion ([Ba/Fe]$\sim -0.63$) typical 
of its iron abundance, degraded to the MIKE resolution. The topmost spectrum is a synthesis
 using stellar parameters similar to Her-2 and a strong [Ba/Fe] deficiency of $-$2 dex.}
\end{figure}
\begin{figure}[tb]
\begin{center}
\includegraphics[angle=0,width=1\hsize]{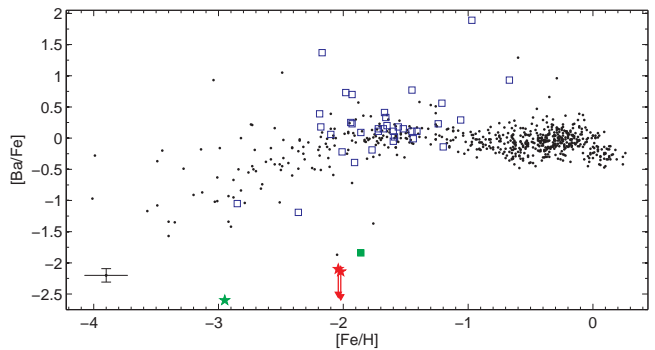}
\end{center}
\caption{Same as Fig.~1 but for Ba abundances.}
\end{figure}

The same holds for every other heavy n-capture element. We do not detect the
usually prominent \ion{Sr}{2} line at 4077\AA\ above noise, nor are 
Y, La, and Eu measurable above the noise, indicating an extraordinary level of depletion.
\section{Inhomogeneous chemical enrichment in Hercules}
The only other red giant in a dSph known to date to have strong enhancements in 
the light elements O, Mg, Na and Al,  and  a complete n-capture deficiency is Dra~119 
(Shetrone et al. 1998; Fulbright et al. 2004). 
Similar to our Her stars, Dra~119 is characterized by no significant occurrence of  elements heavier than Ni.
Fulbright et al. (2004) argue that the chemical patterns seen in Dra~119 
are explicable with SNe II enrichment by high-mass progenitors of at least 20 M$_{\sun}$. 
Theoretical yields (e.g., Woosley \& Weaver 1995) indicate that 
the high ratio of hydrostatic to explosive $\alpha$ elements (O and Mg, relative
to Ca, Ti) found in our Her stars (at [Mg/Ca]= 0.94 and 0.58 dex, respectively), 
requires nucleosynthesis by relatively high mass Type II SNe
($\sim$35 M$_{\odot}$).
Moreover, given the strong depletion of Ba in the two Her stars,  SNe events in this mass regime 
apparently do not produce any significant amount of n-capture elements.

Given the low luminosity of the Her dSph, we consider an incomplete sampling
of the high-mass end of the IMF as the most likely reason for the dominance
of material from $\sim$35 M$_{\odot}$ SNe~II (e.g., Carigi \& Hernandez 2008). 
 Accordingly, we have
performed stochastic chemical evolution experiments to investigate the 
 [O/Mg], [Mg/Ca] and [Mg/Fe] ratios, similar to McWilliam \& Searle (1999).  Our
calculations employed a Miller-Scalo (1979) IMF with slope of $-$2.56 and
element yields from Woosley \& Weaver (1995) for low-metallicity (Z=10$^{-4}$ Z$_{\odot}$) SNe~II 
progenitors in the mass range 12 to 40 M$_{\odot}$.  Our calculations indicate 
that the observed [Mg/Ca] and [Mg/Fe] ratios in Her are found in less than
1\% of systems with 7 to 12 SNe~II events; approximately 10\% of systems possess
the observed [Mg/Ca] and [Mg/Fe] ratios after 1--3, and $\sim$5 SN~II events
respectively.  
About 10\% of systems show the observed [O/Mg] ratios after 11 SNe II;
this reduced to 1\% of systems after 32 SNe II. 
Thus, our experiment suggests that the abundance ratios can only be reasonably obtained in systems
which experienced {\em fewer than $\sim$11 SNe~II} enrichment events, although it is more
likely that only 1--3 events were responsible for the chemical composition 
of our stars.

Seemingly at odds with the idea of small numbers of massive SNe~II is the 
metallicity, at [Fe/H]=$\sim$$-$2.0 dex;  comparable abundance patterns 
in the Galactic halo stars are only found towards the very metal poor regime 
(e.g., McWilliam \& Searle 1999;  Cohen et al. 2007. Dra~119, with 
a comparably high [Mg/Ca] ratio has a very low [Fe/H] of almost $-3$ dex.  
Since Audouze \& Silk 
(1995) argue that most individual SNe produce [Fe/H]$\sim$$-$4.0 dex, we might statistically 
expect that $\sim$100 SNe contributed to the composition of our Her stars.
Other notable abundance anomalies in our Her stars include the extremely low
[Ba/Fe] and enhanced [Co/Fe] ratios.  Remarkably, low [Ba/Fe], high [Co/Fe] and 
low [Cr/Fe] ratios 
are typical of halo stars with [Fe/H]$\sim$$-$3.5 (McWilliam et al. 1995). It is, however, possible, that 
unusual [Co/Cr] abundance ratios may be partly due to NLTE effects. 
Overall, it is as if the processes typical of  the extremely low metallicity 
Galactic halo  occurred in the Her dwarf at [Fe/H]=$-$2.0 dex. 
We note, however, that the high [Mg/Ca] ratios are  inherent to the Her stars and not found 
in the metal poor halo stars.  

It is conceivable that our Her stars formed from SN ejecta  
diluted with a much smaller amount of pristine gas than typically occurred in the
Galactic halo.  
 If the Her stars  in fact  resemble  more the very metal poor 
Galactic halo stars, then their unusual composition could
reasonably be due to one, or at most  a few, SN events. 
This scenario requires that high mass SNe~II produce the enhanced [Co/Fe]
and deficient [Cr/Fe] values seen in our Her stars (typical at extremely
low metallicity).  While this scenario might explain our observed Her abundances,
 it leaves unanswered the origin of the enhanced [Co/Cr] ratios in found in 
 extremely metal-poor stars of the Galactic halo.

A viable interpretation is that our Her stars have an unusually high
abundance of nearly pure population III material, produced and ejected by
the first stars.  These  first stars are thought to be dominated by zero-metallicity high
mass SNe~II (e.g. Yoshida et al. 2008), which would naturally
explain the high [O,Mg/Fe] ratios.  This scenario would also easily explain 
the high [Co/Fe] and low [Cr/Fe] values, if the unusual trends of these two 
ratios at low [Fe/H] are due to population III SN ejecta, as suggested by
McWilliam  (1997).   While this idea removes the
aforementioned difficulties associated with the Co and Cr abundances, the
[Mg/Ca] ratios are unlike halo stars near [Fe/H]=$-$4.
This scenario leads to the question of why the Her dwarf has a concentration
of primordial material $\sim$30 times that of 
the low-metallicity halo.
Perhaps it is related to Her's presumably low 
baryonic mass, which may have formed only one relatively low-mass
population III star that ejected a large amount of population III Fe-peak 
material. 

Whatever the cause, it appears
that stochastic chemical evolution might reasonably explain the unusual
abundance patterns in the Her dwarf galaxy 
(see also Marcolini et al.\ 2006, 2008).
If true,  this would indicate that the
 chemical enrichment  proceeds very differently 
in very low-mass environments (see also Koch et al. 2008),  and that  low
luminosity dwarf galaxies would be useful for directly measuring the abundance
yields of low-metallicity Type II SNe.

\acknowledgments
We gratefully acknowledge  Mark Wilkinson for help with preparing the observations  
and  Andrea Marcolini and Daisuke Kawata  for very  helpful comments on an early version of this Letter.
\end{document}